\documentclass[12pt]{article}
\input epsf
\def\ba{\begin{eqnarray}}
\def\ea{\end{eqnarray}}
\def\be{\begin{equation}}
\def\ee{\end{equation}}

\def\bea{\begin{eqnarray}}
\def\eea{\end{eqnarray}}
\def\bml{\begin{mathletters}}
\def\blea{\begin{mathletters}\begin{eqnarray}}
\def\elea{\end{eqnarray}\end{mathletters}}

\begin{document}
\begin{titlepage}
\begin{flushright}
{\small
NYU-TH-04/08/1}
\end{flushright}
\vskip 0.9cm

\centerline{\Large \bf 
A Note on Tubular Brane Dynamics}
\vspace{1.5cm}

\centerline{\large Jose J. Blanco-Pillado\footnote{E-mail: blanco-pillado@physics.nyu.edu}}
\vskip 0.3cm

\centerline{\em Center for Cosmology and Particle Physics}
\centerline{\em Department of Physics, New York University, New York, 
NY, 10003, USA}

\vskip 1.9cm





\begin{abstract}

{}We present new time dependent solutions for the dynamics of tubular D2-branes.
 We comment on the connection to cosmic string dynamics and explicitly give a 
few simple examples of oscillating and rotating brane configurations.

\end{abstract}

\end{titlepage}


\newpage

\section{Introduction.}

{}Since the discovery of D-branes, one of the main programs in string theory has been
the study of their dynamics. Recently a new type of supersymmetric static solution,
 the supertube, was found in \cite{Mateos-1}. It is basically a bound state of a tubular 
D2 brane and a number of D0 branes and F1 strings dissolved in its worldvolume. 
The physical reason for the tubular brane to remain static can be understood in very simple terms.
The electromagnetic field living on the worldvolume of the D2-brane creates a mechanical effect that
opposes the brane tension. Therefore, there is a limit in which the effective tension for the brane vanishes
and there is no obstacle to have a static configuration. This type of effect was also uncovered
 in the context of superconducting cosmic strings \cite{Witten}. In this case, the presence of a 
current on the string worldsheet plays the same role as the electromagnetic field on the D2-brane. 
This type of static loop solutions are generically found in many superconducting
string models \cite{Davis} but in some cases the equation of state 
is such that allows for the complete solution for the string motion to be found. This is what 
happens for example for the Kaluza-Klein string\footnote{This kind of strings were first discussed 
by Nielsen in the context of a string propagating in an extra-dimensional space \cite{Nielsen}. 
They lead to a 4D superconducting string with an equation of state that has been shown to be
integrable \cite{wiggly1,wiggly2}.} or the chiral string cases \cite{chiral1,chiral2}. In this 
paper we show that this interesting property is also carried over to our tubular brane ansatz.

The paper is organized as follows. In section I we present the equations of motion for a
D2 brane. In section II we give the general solution for these equations within the tubular
brane ansatz. In section III we obtain the known static tubular configurations, supertubes, 
as well as very simple oscillating and rotating cases. Finally we end with some conclusions.

\section{Equations of motion.} 

 The effective action for the worldvolume dynamics of a D2 brane is,

\be
S= - \int d^{3}\xi ~ \sqrt{det~(g_{mn} + F_{mn})},
\ee
where $g_{mn}$ and $F_{mn}$ (with  $m$ and $n = 0, 1, 2$) are the $3\times3$ matrices representing the 
induced metric and the electromagnetic field strength on the worldvolume of the brane. In $2+1$ dimensions
we can always rewrite the action above as:

\be
S= - \int d^{3}\xi ~ \sqrt{det~(g_{mn})}~\sqrt{1+{{1}\over{2}}F^{mn}F_{mn}}.
\ee
The equations of motion for this action can be obtained by taking:

\be
{{\partial}\over{\partial\xi_a}}\left({{\partial {\cal L}}\over{\partial(\partial_{a} x^{\mu})}}\right)=0,
\ee
and also,

\be
{{\partial}\over{\partial\xi_{a}}}\left({{\partial {\cal L}}\over{\partial(\partial_{a} A^{b})}}\right)=0,
\ee
which give respectively,

\be \label{string-equation}
\partial_{a}\left(\sqrt{g}\sqrt{1+{{1}\over{2}}F^{mn}F_{mn}} \left(g^{ab} - {{F^{al} F^{bk}g_{k l}}\over{1+{{1}\over{2}}F^{mn}F_{mn}}}\right)\partial_{b} x^{\mu}\right)=0,
\ee
and

\be \label{A-equation}
\partial_{a}\left({{\sqrt{g}}\over{\sqrt{1+{{1}\over{2}}F^{mn}F_{mn}}}} F^{ab}\right)=0.
\ee

\section{Solutions.}

In order to look for solutions of these equations, we will assume that the D2-brane has a tubular geometry 
so we can parametrize its worldvolume in the following way,

\be
{\bf x}= {\bf x_{\perp}} (\xi_{0},\xi_{1}) + \xi_{2} {\bf z}.
\ee

On the other hand, we restrict ourselves to the case with only electric field in the $\xi_2$-direction but an arbitrary magnetic field
pointing perpendicularly to the surface of the tube. In this ansatz the induced metric and the electromagnetic field strength
look like,

\be
g_{mn}= \left(\begin{array}{ccc}
g_{00} & g_{01} & 0 \\
g_{01} & g_{11} & 0 \\
0      & 0      & -1 
\end{array}
\right)=
\left(\begin{array}{ccc}
\gamma_{00} & \gamma_{01} & 0 \\
\gamma_{01} & \gamma_{11} & 0 \\
0      & 0      & -1 
\end{array}
\right)\,,
\ee
and,
\be
F_{mn}= \left(\begin{array}{ccc}
0 & 0 & F_{02} \\
0 & 0 & F_{12} \\
-F_{02}  & -F_{12} & 0 
\end{array}
\right)\,.
\ee

We can now write the field strength in (2+1) dimensions in terms of a scalar field, $\theta$, in the following
way,

\be\label{duality}
F^{ab}= {{\sqrt{1+{{1}\over{2}}F^{mn}F_{mn}}}\over{\sqrt{g}}} \epsilon^{abc} \partial_c \theta,
\ee
where $\epsilon^{abc}$ is just the Minkowski space Levi-Civita tensor. It is clear that with this parametrization equation
(\ref{A-equation}) is fulfilled. On the other hand, substituting the form of the field strength in equation (\ref{string-equation})
we arrive at,

\be
\partial_{a}\left({{\sqrt{g}}\over{\sqrt{1 - g^{cd} \theta_{,c} \theta_{,d}}}} \left(g^{ab} - {{g_{kl} \epsilon^{alm}  \epsilon^{bkn} 
 \theta_{,m} \theta_{,n}}\over{g}} \right)\partial_{b} x^{\mu}\right)=0.
\ee

Also, since nothing depends on the coordinate $\xi_2$, the only relevant equations for our tubular brane ansatz are,

\be\label{brane-motion}
\partial_{a}\left( {\cal J}^{ab}\partial_{b} x_{\perp}^{\mu}\right)=0
\ee
with
\be
{\cal J}^{ab}={{\sqrt{-\gamma}}\over{\sqrt{1 - \gamma^{cd} \theta_{,c} \theta_{,d}}}}\left(\gamma^{ab}-{{\epsilon^{a2m}  \epsilon^{b2n} \partial_m \theta \partial_n \theta}\over{\gamma}}\right).
\ee

Which we can write explicitly as:
\be
{\cal J}^{ab}= {{1}\over{\sqrt{-\gamma} \sqrt{1 - \gamma^{cd} \theta_{,c} \theta_{,d}}}} \left(\begin{array}{cc}
\gamma_{11}-(\theta')^2 & -\gamma_{01}+(\dot \theta \theta') \\
 -\gamma_{01}+(\dot \theta \theta')& \gamma_{00}-(\dot \theta)^2
\end{array}
\right).
\ee

At this point it is important to remember that we still have the freedom to fix the gauge in the subspace transverse to the axis of symmetry of the tube. Looking at 
equation (\ref{brane-motion}) it seems natural to impose the gauge that makes ${\cal J}^{ab}= \eta^{ab}$, where by $\eta_{ab}$ we denote the flat 
Minkowski metric in 1+1 dimensions. In this gauge the equation for the transverse motion of the brane is just the wave equation, so we can easily
write the most general solution as:

\be\label{brane-motion-solution}
{\bf x}= {1\over{2}}({\bf a_{\perp}} (\sigma-\tau)+ {\bf b_{\perp}}(\sigma + \tau)) + \xi_{2} {\bf z},
\ee
where we have set $x^0=\xi_0=\tau$ and $\xi_1=\sigma$. Imposing the constraints, we arrive at the conclusion 
that the solution for the scalar field $\theta$ has to be of the form,

\be\label{theta-solution}
\theta(\tau,\sigma)=  {1\over{2}} (f (\sigma-\tau) + g (\sigma+\tau)).
\ee

With this parametrization for the solution the rest of the constraints give the following relation,

\be\label{constraint-1}
|{\bf a'}|^2 + (g')^2 =1
\ee

\be\label{constraint-2}
|{\bf b'}|^2 + (f')^2 = 1
\ee

In order to obtain a closed tubular brane solution we have to impose the correct periodicity on the functions
{\bf a }, {\bf b}, f and g and this in turn means that the solutions would be periodic. 

We see from eqns. (\ref{brane-motion-solution},\ref{theta-solution}) and (\ref{constraint-1},\ref{constraint-2})
 that we can interpret these solutions for the transverse brane motion as a tubular brane
propagating in one more dimension, where the field $\theta$ parametrizes the position of the string along the
extra dimension. In fact this interpretation clarifies the origin of the relation (\ref{duality}) as a duality transformation
between the tubular D2-brane solution with an electromagnetic field on its worldvolume and a M2-brane propagating
in one extra compact dimension \cite{Super-P}. Similar duality transformations have been used to map the supertube solution
into several other objects within M-theory \cite{Helix,Mateos-1,Supercurve,Ribbons}. This is of course also possible with the solutions presented here.

\section{A few examples.}

\subsection{Supertubes.}

It is straightforward to use the equations presented in the previous sections to get simple solutions of the 
equations of motion for tubular branes. Perhaps the most important ones are the {\it supertubes}. Let us see how they can
appear in our description. It is clear that whenever you have a solution with $(g')^2 =1$ or  $(f')^2 =1$ the transverse position of the
brane would not change with time. It may seem to be time dependent but actually the points in space that the brane occupies are always the same.
Also it is clear that since the only constraint on {\bf b} or {\bf a} in this case is that their modulus is constant, the shape of the brane can be 
arbitrary. These are also the main characteristics of the supertube solutions previously studied in the literature. 

The corresponding type of solutions in the superconducting string models are the so called {\it vortons} \cite{Davis} that have also been shown to exist for
an arbitrary shaped loop. (See, for example \cite{chiral2}).

\subsection{Oscillating solutions.}

On the other hand, the solutions presented in this paper are not in general static, after all, we know that 
in the absence of any field strength on the worldvolume we would expect the brane to contract under its own tension and collapse. 

As we mentioned in the introduction, the intuitive reason for the existence of the static solutions is the possibility
that the electromagnetic backreaction on the brane cancels its tension, so a natural question to ask is what happens when this cancellation 
is not complete. Here we describe the simplest possible solutions that are an intermediate step between the collapsing branes and the supertubes.
Let's take the following solution,

\bea
{\bf a} &=& A \sin(\sigma-\tau)~{\bf x} + A\cos(\sigma-\tau)~{\bf y} \\
{\bf b} &=& B\sin(\sigma+\tau)~{\bf x} + B\cos(\sigma+\tau)~{\bf y} \\
\theta &=& {1\over2}\left(\sqrt{1-A^2} (\sigma-\tau) + \sqrt{1-B^2} (\sigma+\tau)\right).
\eea

Taking for simplicity $A=1$ we can find, depending on the value of $B$, the three different behaviors 
we commented on previously; the supertube which corresponds to the case $B=0$, the collapsing brane 
which has $B=1$ and the oscillating circular brane for any value of $B$ within the range, $0< B< 1$.

\subsection{Rigidly rotating solutions.}

A minor change on the previous parametrization yields brane configurations that are rigidly rotating on a plane. These are the analog 
solutions of the well-known Burden family of rotating cosmic string solutions \cite{Burden},

\bea
{\bf a} &=& {A\over N}\left(\sin\left[N(\sigma-\tau)\right]~{\bf x}+\cos\left[N(\sigma-\tau)\right]~{\bf y}\right) \\
{\bf b} &=& {B\over M}\left(\sin\left[M(\sigma+\tau)\right]~{\bf x}+\cos\left[M(\sigma+\tau)\right]~{\bf y}\right) \\
\theta &=& {1\over2}\left(\sqrt{1-A^2} (\sigma-\tau) + \sqrt{1-B^2} (\sigma+\tau)\right).
\eea

We show in Fig. (1) two different examples of this family of solutions with the following parameters, $A=1, B={1\over2}, N=1 and M=\pm 5$.
The major effect of the electromagnetic backreaction 
on the brane motion is the smoothing out of the high curvature regions and a general slow down of the average brane motion. 
These are in fact two general features that can be found in all the cosmic string scenarios with some backreaction mechanism \cite{chiral2, chiral-cusps},
 and it is therefore also expected to be a result in this case.

There are of course many other solutions apart from the simple cases we presented here that would be simultaneously rotating and
oscillating. In fact, it is possible to use the methods developed in \cite{Ken_and_Xavier} to generate solutions that we could
readily interpret as solutions of the tubular brane ansatz.

\begin{figure}
\centering\leavevmode \epsfysize=6cm \epsfbox{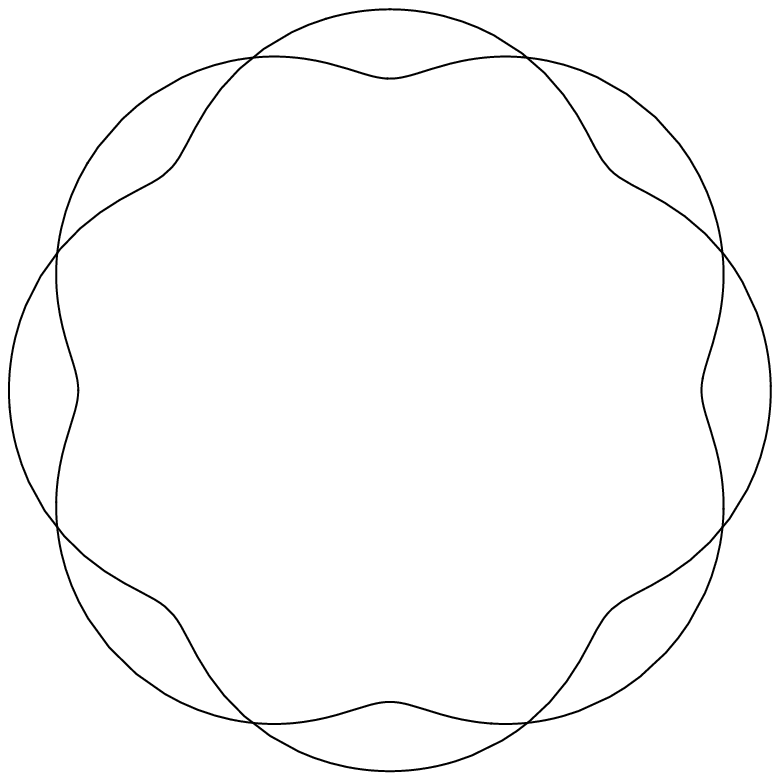}
\epsfysize=6cm \epsfbox{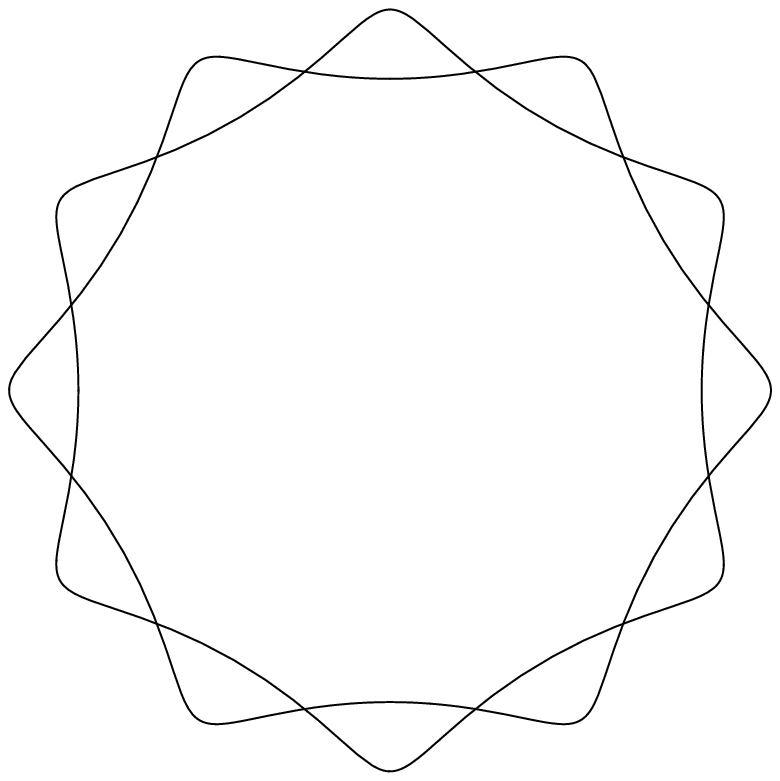}
\caption[Fig 1] {Snapshots at two different times in their evolution of the cross section of the rigidly rotating tubular brane solutions described in the main text.}
\end{figure}

\section{Conclusions.}

We have shown how to obtain new solutions for the dynamics of D2-branes with a tubular geometry.
These are in general time dependent solutions which can be reduced in a particular limit to the
previously known static solutions, the supertubes. We illustrate the effect of the electromagnetic
field backreaction on the brane motion by explicitly showing some simple time dependent solutions 
of oscillating and rotating branes. We speculate with the possibility that some of these solutions
could be useful to study excited configurations of the supertube solutions.

\section{Acknowledgments.}

{} I would like to thank Roberto Emparan, Jaume Garriga, Marta Gomez-Reino, Alberto Iglesias, David Mateos, Ken Olum, 
Josep Taron and Alexander Vilenkin for useful discussions. This work is supported by the James Arthur Fellowship
at NYU.

\end{document}